\documentclass[useAMS,usenatbib]{mn2e}
\usepackage{graphicx}
\usepackage{psfrag}

\title[TeV lightcurve of PSR B1259-63/SS2883]
{TeV lightcurve of PSR B1259-63/SS2883}
\author[Khangulyan,  Hnatic, Aharonian \& Bogovalov]
       {D. Khangulyan${}^1$\thanks{E-mail:dmitry.khangulyan@mpi-hd.mpg.de }, S. Hnatic${}^1$, F. Aharonian${}^1$ \& S.Bogovalov${}^2$ \\
        ${}^1$ Max Planck Institut f\"ur Kernphysik, Heidelberg 69117, Germany\\
	${}^2$ Moscow Engineering Physics Institute (State University), Moscow 115409, Russia}

\begin{document}

\date{Accepted. Received; in original form}

\pagerange{\pageref{firstpage}--\pageref{lastpage}} \pubyear{}

\maketitle

\label{firstpage}

\begin{abstract}
The inverse Compton (IC) scattering of ultrarelativistic electrons accelerated at  the pulsar wind termination shock is generally believed to be responsible for TeV gamma-ray signal recently reported from the  binary system PSR B1259-63/SS2883. While this process  can explain the energy spectrum of the observed  TeV  emission,  the  gamma-ray fluxes detected by HESS at different epochs do not agree with the published theo\-retical predictions of the TeV lightcurve. { The main objective of this paper is to show that the HESS results can be explained, under certain reasonable  assumptions concerning the cooling of relativistic electrons, by  inverse Compton scenarios of gamma-ray production in PSR B1259-63.} In this paper we study evolution of the energy spectra of relativistic electrons under different assumptions about the acceleration and energy-loss rates of electrons, and the impact of these processes on  the lightcurve of IC gamma-rays.  We demonstrate that the observed TeV lightcurve can be explained (i) by adiabatic losses which dominate over the entire trajectory of the pulsar with a significant increase towards the periastron, or (ii) by the "early"  (sub-TeV) cutoffs in the energy spectra of electrons due to the enhanced rate of  Compton losses close to the periastron. { The first four data points obtained just after periastron comprise an exception - possibly due to interaction with the Be star disk, which introduces additional physics not included in the presented model.} The calculated spectral and temporal characteristics of the TeV radiation provide conclusive tests to distinguish between these two working hypotheses. The Compton { deceleration} of the  electron-positron pulsar wind contributes to the decrease  of the nonthermal power released in the accelerated electrons after the wind termination, and thus to the reduction of the IC and synchrotron components  of radiation close to the periastron. Although this effect alone cannot explain the observed TeV and X-ray lightcurves, the Comptonization of the cold ultrarelativistic wind leads to the formation of gamma-radiation  with a specific line-type energy spectrum. { While  the HESS data already constrain the Lorentz factor of the wind, $\Gamma \le 10^6$ (for the most likely orbit inclination angle $i=35^\circ$, { and assuming an isotropic pulsar wind})}, future observations of this object with GLAST should allow a deep probe of the wind Lorentz factor in the range between $10^4$ and   $10^6$. 
\end{abstract}


\section{Introduction}
PSR B1259-63/SS 2883 -- a binary system consisting of a 47ms pulsar orbiting around  a luminous Be star \citep{johnston92} -- is a unique high energy laboratory for the study of nonthermal processes  related to the ultrarelativistic pulsar winds. X-ray and gamma-ray  emission components are  expected from  this object due to the radiative (synchrotron and inverse Compton) cooling of relativistic electrons accelerated by the wind  termination shock  \citep{tavani97,kirk99}. Generally,  the particle acceleration  in this complex system can be treated as a scaled-down in space and time ("\textit{compact and fast}") realization  of the  current  paradigm  of Pulsar Wind Nebulae (PWN)  which suggests  that the interaction of the ultrarelativistic pulsar wind with  surrounding medium  leads to the formation of a relativistic  standing shock \citep{rees74,kennel84,harding90,tavani97}.   
In the case of strong and young  pulsars,  the shock-accelerated multi-TeV  electrons  should give rise to observable  X-ray (synchrotron)  and  TeV (inverse Compton)  nebulae with typical linear size $\sim 0.1 -10$ pc. The unambiguous association of some of the recently discovered  extended TeV gamma-ray sources  with several  distinct synchrotron X-ray PWNs generally
supports this scenario of formation of nonthermal nebulae around the pulsars.  

In the binary system  PSR B1259-63/SS 2883  one should expect a similar mechanism of conversion of the major fraction of the  rotational energy of  the pulsar to ultrarelativistic electrons through formation and termination of the cold electron-positron 
wind. On the other hand,  in such systems the magnetohydrodynamic (MHD), acceleration and radiation processes proceed under essentially different conditions compared to the PWN around isolated pulsars.  In particular, due to the high pressure of the ambient medium caused 
by the outflow from the companion star,  the pulsar wind terminates quite close to the pulsar, $R \leq 10^{12}$ cm.   Consequently,  in such systems particle acceleration occurs  at presence of much stronger magnetic  field ($B \sim 0.1 -1$ G), and under illumination of intense radiation from the normal star with a density 
\begin{equation}
w_{\rm ph}={L_{\rm star}\over4\pi c R^2}\sim
0.9 \left(L_{\rm star}\over3.3\cdot10^{37}{\rm erg/s}\right) \left(R \over10^{13}{\rm cm}\right)^{-2} \ {\rm erg/cm^3}\,,\label{energy_phot}
\end{equation}
where $L_{\rm star}$ is the luminosity of the Be star and  $R$ is the distance between the acceleration site and the Be star
. The discussed ranges of the temperature and the luminosity of the star SS2883 vary within  $T\simeq 2.3\cdot10^4-2.7\cdot10^4 {\rm K}$ and \mbox{$L_{\rm star}\simeq3.3\cdot10^{37}-2.2\cdot10^{38}{\rm erg/s}$}. This implies that both the acceleration and radiative cooling timescales of  TeV electrons are  of order of  hours, i.e. comparable or shorter than the typical dynamical timescales characterizing the system. This allows a unique "on-line watch" of the extremely complex MHD processes of 
creation and termination of the  ultrarelativistic pulsar wind and the subsequent
particle acceleration,  through the study of spectral and temporal characteristics of  high energy  gamma-radiation of the system. The discovery of TeV gamma-radiation from PSR B1259-63/SS2883 by HESS collaboration \citep{aharonian05a} provides  the first 
unambiguous evidence  of  particle acceleration in such systems to TeV energies. 

Remarkably, in spite of the complexity of the binary system PSR B1259-63/SS2883, one may calculate with  quite high  precision the spectral and temporal features of gamma-radiation based on only a few model assumptions concerning, in particular, the magnetic field, the acceleration  rate as well as the  non-radiative (adiabatic or escape) losses of electrons.  The { observational} uncertainties involved in these calculations are mainly related to the luminosity of the optical companion SS 2883. The basic free parameters used in calculations are the  rates of particle acceleration  and the nonradiative losses caused by adiabatic expansion and escape of electrons.

\begin{figure}
\includegraphics[width=5cm,angle=270]{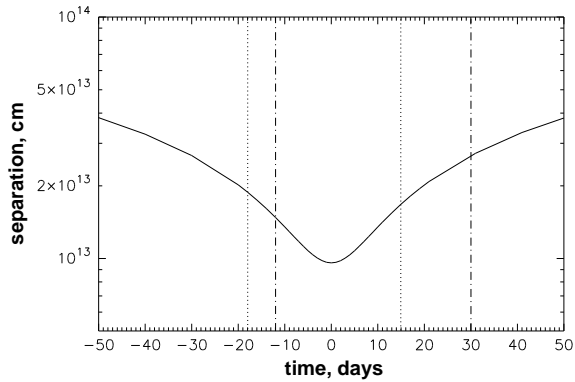}
\caption{Separation between the pulsar and the star versus 
time (0 corresponds to the periastron passage).  The vertical 
dotted lines correspond to the location of interaction of the 
pulsar wind with the stellar disk based on the observations of the  
eclipse of the pulsed radio emission  \citep{johnston05}. 
The vertical dashed-dotted lines 
correspond to the disk location 
suggested by Chernyakova et al. (2006).}
\label{separation}
\end{figure}

The orbital elements of the system are well known. { The orbit is quite eccentric ($e=0.87$) with the minimum and maximum distances between the pulsar and the star $D_{\rm 0}=9.6\times 10^{12}{\rm cm}$ and $D_{\rm a}=1.4\times 10^{14}{\rm cm}$ at the periastron and the apastron, respectively; the inclination of orbit is $i\simeq35^\circ$ \citep{johnston92}}. The separation between the pulsar and the optical companion versus the epoch is shown  in Fig.\ref{separation}. In the same figure we show the possible locations of pulsar passage through the stellar disk as discussed by  \cite{johnston05} and \cite{chernyakova06}.  In the first paper the disk location is determined by the time of disappearance of the pulsed radio emission which is  explained by absorption of the  pulsar emission in the stellar disk. However, as long as the physics of interaction 
of the pulsar winds with the stellar disk is not  firmly established,  alternative models are not excluded. For example, \cite{chernyakova06} noticed that the maxima of lightcurves of \textit{ nonpulsed} radio, X-ray and TeV gamma-ray are quite close to each other, and proposed that the increase of the nonthermal energy release happens  
when the pulsar crosses the disk. If so, the gamma-ray emission could be result of hadronic interactions \citep{kawachi04,chernyakova06}. This hypothesis implies, however, a different location of the disk compared to the one derived from the 
eclipse of the pulsed radio emission \citep{johnston05,bogomazov05}, and therefore requires an independent confirmation based on a stronger evidence  of correlation of radio, X-ray and TeV  fluxes, as well as detailed study of the reasons of such correlation. 

Meanwhile, the  inverse Compton scattering remains the most plausible gamma-ray production
mechanism \citep{tavani97,kirk99,aharonian05a}. In this paper we present detailed numerical studies of the the spectral and temporal characteristics of TeV gamma-ray emission 
within the framework of the IC model of TeV gamma-rays.
In this context we  adopt the position of the disk as it is derived from the eclipse of pulsed radio emission \citep{johnston05}.

The position of the shock wave is determined by interaction of the pulsar wind with the {stellar wind},  therefore the distance to the shock is a function of time.  For
the magnetic field lines frozen into the pulsar wind, one has $B\propto r_{\rm sh}^{-1}$. 
It is also expected that $r_{\rm sh}\propto D$ \citep{kirk99}, thus $w_{\rm B}\propto D^{-2}$. { As long as the mass flux density from a Be star is expected to be significantly higher than one from the pulsar \citep{waters}, we adopt} $r_{\rm sh} \ll D$, and the target photon density at the site of electron acceleration and radiation  $w_{\rm ph}={L_{\rm star}/(4\pi c) D^{-2}}$. Thus the synchrotron and IC radiation timescales  have similar dependencies on the separation distance $D$, namely $t_{\rm syn},t_{\rm IC}\propto D^2$.

{ The magnetic field strength in the pulsar wind at distance $r$ from the pulsar can be estimated as the following
\begin{equation}
B=\sqrt{\sigma L_{\rm sd}\over(1+\sigma)cr^2}\,,
\end{equation}
where $L_{\rm sd}$ is spindown luminosity of the pulsar; and $\sigma$ is the ratio between the Poynting and kinetic energy flux in the pulsar wind. While the spindown luminosity is entirely known and is measured to be $L_{\rm sd}=8.3\cdot10^{35}$~erg/s; the $\sigma$ parameter is not constrained either theoretically or observationally. In what follows we assume the magnetic field at the termination shock to be $B\simeq 0.1-1 {\rm G}$ around the periastron epoch. This magnetic field strength corresponds to a value of $\sigma=0.04-4\cdot10^{-4}$ (for the distance between the pulsar and the termination shock $r_{\rm sh}=10^{12}$~cm), which is consistent with the value $\sigma=0.02$ adopted by \cite{tavani97}. As long as the $\sigma$-parameter is rather small, we do not study some minor effects, e.g. magnetic field amplification on the termination shock, or  possible impact of magnetic field adjustment on the particle flux and the Lorentz factor of the pulsar wind.

For the expected magnetic field strength, the corresponding energy density is $B^2/8\pi \sim  10^{-3} - 10^{-1} {\rm erg/cm^3}$.}  
The energy density of the photon field significantly exceeds this value (see Eq. (\ref{energy_phot})). Thus the radiation is formed  in an environment dominated by radiation. Since the temperature of the starlight is about 2 eV, the inverse Compton scattering proceeds in a regime with distinct  features related to the transition from the Thomson to Klein-Nishina limits, depending on the scattering angle (i.e. location of the pulsar in the orbit) and the { electron energy} \citep{khangulyan04}.

The TeV gamma-ray lightcurve  of binary PSR B1259-63/SS2883 shows \citep{aharonian05a}
a tendency for a minimum flux at the epoch close to the periastron passage, as well as a maximum observed 20 days after the periastron. Although the available TeV data do not allow robust conclusions about the lightcurve before the periastron, there is evidence of a time variable flux which indicates the existence of the second maximum $>$18 days  before the periastron. { While the light curve reported by HESS needs independent confirmation by future measurements, throughout this paper we assume that the TeV flux decreases towards the periastron, as stated by the HESS collaboration \citep{aharonian05a}.} Interestingly, the X-ray observations show a similar behavior \citep{tavani96,chernyakova06}. 

Below we discuss 3 different possible scenarios which could explain the drop of the TeV gamma-ray luminosity close to periastron: (i) nonradiative (adiabatic or escape) 
losses of electrons; (ii) "early"  (sub-TeV) cutoffs in the energy spectra of shock-accelerated electrons due to the increase of the rate of  Compton losses, and  (iii) decrease of the kinetic energy of the pulsar wind before its termination due to the Comptonization of electrons in the cold ultrarelativistic wind. Note that generally in the X-ray binaries with luminous companion stars the photon-photon absorption may
have a strong impact in the formation of the  TeV gamma-ray lightcurves. However, in the 
case of  PSR B1259-63/SS2883 the absorption effect appears to be not significant \citep{kirk99,dubus05}.  

\section{The electron distribution function}
Formally, the radiation seen by an observer is contributed by electrons of different ages, i.e. by electrons from different locations of the pulsar during its orbiting. However, since the radiative cooling time of high energy electrons ($E \geq $ 100 GeV) is quite short (see below), we effectively see the radiation components 
from a localized part of the orbit with  homogeneous physical conditions. This allows us to reduce the treatment of time-evolution of energy distribution of electrons to the well-known equation (see e.g. \cite{ginzburg64})
\begin{equation}
{\partial n\left(t,\gamma\right)\over\partial t}+{\partial \dot{\gamma} n\left(t,\gamma\right)\over\partial \gamma}+{n\left(t,\gamma\right)\over T_{\rm esc}}=Q(t,\gamma)\,,
\label{Ginz_equation}
\end{equation}
where $\dot{\gamma}=\dot{\gamma}_{\rm ic}+\dot{\gamma}_{\rm synch}+\dot{\gamma}_{\rm ad}$;$\dot{\gamma}_{\rm ic}$, $\dot{\gamma}_{\rm synch}$, $\dot{\gamma}_{\rm ad}$ are electron energy loss rates { (IC, synchrotron and adiabatic, respectively)} and $Q(t,\gamma)$ is the acceleration rate. The applicability of the Eq.(\ref{Ginz_equation}) to the case of IC losses in the Klein-Nishina regime was shown e.g. in \cite{khangulyan04}. The solution of this equation has the following form %
\begin{equation}
n(t,\gamma)={1\over|\dot{\gamma}|}\int\limits_{\gamma}^{\gamma_{\rm eff}} Q(t-\tau,\gamma'){\rm e}^{-\tau({\gamma,\gamma'})/T_{\rm esc}}{\rm d}\gamma'\,,\\
\label{Ginz_solution}
\label{solution0}
\end{equation}
%
where $Q(t,\gamma)$ is the acceleration rate at the given epoch; and { $\gamma_{\rm eff}$ is implicitly defined by the following equation:}
\begin{equation}
t=\int\limits_{\gamma}^{\gamma_{\rm eff}}{{\rm d}\gamma'\over|\dot{\gamma'}|}\,.
\end{equation}
{ And for $\tau(\gamma,\gamma')$ one has}
\begin{equation}
\tau(\gamma,\gamma')=\int\limits_{\gamma}^{\gamma'}{{\rm d}\gamma''\over|\dot{\gamma''}|}\,.
\end{equation}

Since the cooling time of electrons is much shorter than the characteristic dynamic times of the system,
we can use the steady-state distribution function of electrons at given epoch $t$ (or at given position of the pulsar in the orbit),
\begin{equation}
n(t,\gamma)={1\over|\dot{\gamma}|}\int\limits_{\gamma}^{\gamma_{\rm max}} Q(t,\gamma'){\rm e}^{-\tau({\gamma,\gamma'})/T_{\rm esc}}{\rm d}\gamma'\,,\\
\label{solution}
\end{equation}
where $\gamma_{\rm max}$ is the maximum Lorentz factor of injected particles.

It should be noted that at low energies this solution may have a limited applicability  because of long radiative cooling time of electrons. The minimum energy of electrons for which the solution  Eq.(\ref{solution}) remains correct, is determined 
from the condition for the cooling time: $t_{\rm cooling}< t_{\rm dyn}$ where $t_{\rm dyn} \sim 1\,{\rm day}$ (the time during which the distance between the pulsar and the optical star, and  other principal parameters change less than $10 \%$, even at the epochs close to the periastron). This condition gives  
\begin{equation}
E \geq  100 \left({w_{\rm ph}+w_{\rm B}\over {1\rm \,erg\,cm^{-3}}}\right)^{-1}\,{\rm MeV}\,.
\label{minenergy}
\end{equation}

{While very high energy electrons "die" due to radiative losses inside the acceleration region and cannot effectively escape, low energy electrons can escape from the source. The maximum energy of electrons which escape from the source is determined by the characteristic escape time,
}
%
\begin{equation}
E<10 \left({w_{\rm ph}+w_{\rm B}\over {1\rm \,erg\,cm^{-3}}}\right)^{-1}\left({T_{\rm esc}\over 10^3{\rm s}}\right)^{-1}\,{\rm GeV}\,.
\label{Eesc}
\end{equation}

These electrons form a quasi-stationary halo around the binary system which can contribute to the overall IC radiation of the source. { This component is to be formed in the Thomson regime ($E_\mathrm{\gamma}\sim\epsilon_{\rm ph}\gamma_\mathrm{e}^2=(3 kT)\gamma_e^2$) and  taking into account Eq.(\ref{Eesc}) one obtains a typical energy of the halo radiation,}
\begin{equation}
E_{\gamma, halo} \sim  
2.5 \left ({T_{\rm esc}\over 10^3{\rm s}}\right)^{-2}\,{\rm GeV}\,.
\label{halo}
\end{equation}

This radiation can be detected by GLAST as a quiescent component. Note however that it can  be significantly suppressed in the case of a \textit{low-energy cutoff} in the acceleration spectrum of electrons as is often assumed for PWN  in general, and for 
this binary system, in particular (see e.g. \cite{kirk99}).   
 
Below we assume a power-law distribution  for the accelerated { electrons} with an exponential \textit{high energy cutoff}, $E_{\rm e,max}$: 
\begin{equation}
Q(t,\gamma)= A \gamma^{-\alpha}{\rm \bf exp}
\left[-\gamma\,mc^2/E_{\rm e,max}\right]\,,
\label{injection}
\end{equation}
where A is the normalization coefficient { related to the fraction of pulsar wind particles turned into a high-energy power-law distribution}. The cutoff energy $E_{\rm e,max}$ is
determined from the balance between the acceleration and energy loss rates, therefore it is a function of time. Generally, the total acceleration power determined by the parameter $A$ is also time-dependent.  

If the radiative cooling times of electrons are smaller than the adiabatic and escape losses, the radiation of electrons proceeds in the saturation regime, i.e. the source works as a calorimeter. In this regime the energy of accelerated electrons is radiated away due to synchrotron and IC processes:
\begin{equation}
L_{\rm syn}={t_{\rm syn}^{-1}\over t_{\rm syn}^{-1}+t_{\rm IC}^{-1}}L_{\rm injection}\,,
\end{equation} 
\begin{equation}
L_{\rm IC}={t_{\rm IC}^{-1}\over t_{\rm syn}^{-1}+t_{\rm IC}^{-1}}L_{\rm injection}\,f(\theta)\,.
\end{equation}
The function $f(\theta)$ is determined by the angular anisotropy of IC radiation, where  $\theta$ is the angle between the line of sight and the line connecting the pulsar and the Be star. In principle, the function $f$ can vary significantly, but in the case of PSR B1259-63/SS2883 the variation of this function does not exceed two. 

\begin{figure}
\includegraphics[width=6cm,angle=270]{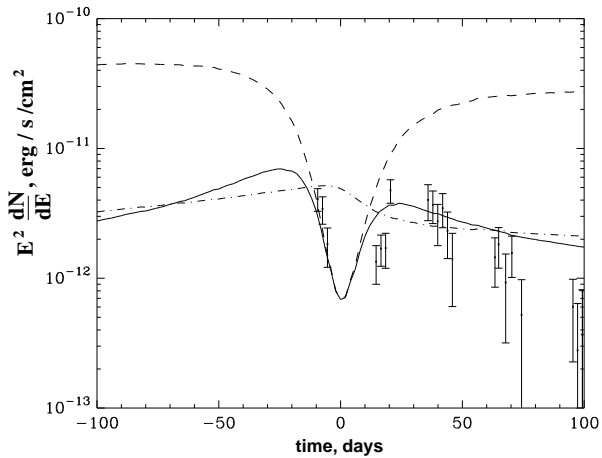}
\caption{Inverse Compton gamma-ray lightcurves calculated for 1 TeV photons. The curves correspond to three  different models.  \textit{(i)} The dashed-dotted line corresponds to 
the constant injection power and energy cutoff in the spectrum of electrons.   
\textit{(ii)} The dashed line corresponds to the case similar to (i), but assuming time-dependent energy cutoff (see Sec.\ref{maxenergy})). The parameter 
$\eta$ which characterizes acceleration rate is fixed at the level $\eta=7\cdot10^3$; the magnetic field varies with the separation distance as 
$B(D) \propto 1/D$ with  $B_0=0.05\ {\rm G}$ at periastron. 
\textit{(iii)} The solid line corresponds to the case similar to 
(ii) with an additional assumption about escape of particles
with escape time $T_{\rm esc}=5\cdot10^3\, {\rm s}$). 
The experimental points are from the HESS observations at energy 1 TeV
\citep{aharonian05a}.}
\label{lightcurve1}
\end{figure}

The lightcurve of 1 TeV gamma-rays produced by electrons with injection spectrum given 
by Eq.(\ref{injection}) with \textit{time-independent} parameters $A$ and $E_{\rm e,max}$  is shown in Fig.\ref{lightcurve1} by dash-dotted line. It is important to note that 
the almost time-independence of this curve, with a weak maximum just before the periastron, is the result of the anisotropy of IC scattering.  Such a lightcurve which implies almost constant flux (within a factor of two) over the entire orbital period, is in obvious conflict with the HESS observations which show noticeable reduction of the flux 
towards the periastron \citep{aharonian05a}. To achieve such a behavior of the lightcurve, { assuming a constant electron injection,} one should introduce additional energy losses. This cannot be achieved by increasing the magnetic field, because it would lead to an increase of the synchrotron 
flux close to the periastron, in contrast to X-ray observations \citep{tavani97,chernyakova06}. This implies that one needs to introduce additional 
"invisible", i.e. nonradiative energy losses. This case is discussed in  \mbox{Section \ref{nonrad}}.  Alternatively, one may reduce the flux both of synchrotron X-rays and of IC gamma-rays assuming a tendency of decrease of cutoff energy in the  spectrum of accelerated electrons $E_{\rm e,max}$, or assuming reduction of the total power of accelerated electrons close to the periastron. These two cases are discussed in \mbox{Sections \ref{maxenergy}
and \ref{wind}}, respectively.

\section{Nonradiative  losses}\label{nonrad}
The interaction of the pulsar wind with the ambient medium 
results in  a complex shock wave structure where adiabatic 
and escape losses may play a dominant role.  
Indeed,  the escape time $T_{\rm esc}$ can be as short as 
\begin{equation}
T_{\rm esc}={D\over (c/3) }\simeq{\left\{10^3\, {\rm s \,(for\, periastron)}\atop 10^4\,{\rm s\,(for\,apastron)}\right.}\,,\label{esctime}
\end{equation}
i.e. comparable or shorter (depending on energy) 
than the radiative cooling time of electrons.
The adiabatic losses ($t \sim R/v$, 
where $R$ is the characteristic radius 
and $v$ is the speed of expansion of the the emission region) 
can be even faster since  $R \ll D$ and the source can expand
relativistically. { We note, that although under certain conditions the particle escape and the adiabatical loss times may be rather connected, in general these two timescales are indeed different. For example, in the case of fast diffusion, electrons diffuse away from the source, without suffering any adiabatical losses.} 

\begin{figure}
\includegraphics[width=8cm]{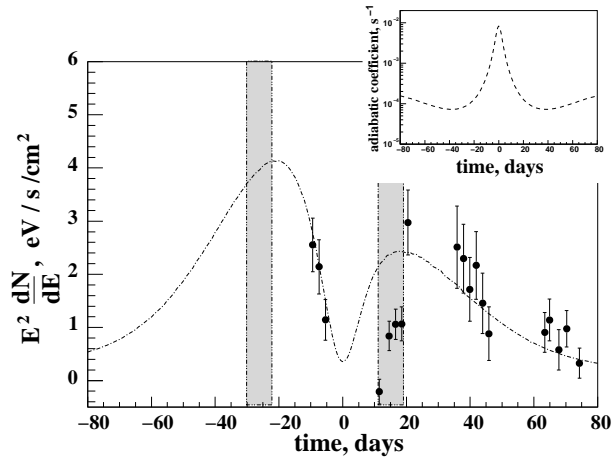}
\caption{Main panel: The lightcurve of 1 TeV gamma-rays detected by HESS from PSR B1259-63/SS2883 \citep{aharonian05a}. A reference lightcurve adopted for derivation of the time profile of nonradiative energy losses of electrons is also shown. The two vertical gray zones correspond to the position of the stellar disk. The somewhat lower flux of gamma-rays at 
$t \sim 15$ days after the periastron can be associated with
the enhanced losses in the disk, and thus may cause more irregular 
profile of energy losses. Small panel: The reconstructed time profile of adiabatic energy loss rate derived for the reference TeV lightcurve shown in the main panel (see the text).}  
\label{ad_light_curve}
\end{figure}

Both adiabatic and escape timescales cannot be calculated 
from the first principles given the complexity of the system\footnote{ Concerning this complicated issue, we would like to note a possibility to get this information by a direct numerical simulation \citep{bogovalov07}.}. 
Instead, the characteristic timescales of nonradiative losses 
can be derived phenomenologically, namely 
from the observed gamma-ray lightcurve.  As discussed above, 
in order to explain the reduction of both X-ray and 
TeV gamma-ray fluxes 
close to the periastron, one should increase 
the rate of nonradiative losses. The reported TeV gamma-ray observations
are quite sparse and unfortunately allow a broad 
range of lightcurves.  In Fig. \ref{ad_light_curve} we show
an example of a lightcurve which matches the HESS data. 
The time-profile of the rate of nonradiative losses 
derived from the lightcurve along with the additional assumption
that  10\% of pulsar spindown luminosity 
is converted (through the termination shock) 
into relativistic electrons with constant (for all epochs) 
injection rate  and energy spectrum given by 
Eq.(\ref{injection}) with $E_{\rm e,max}=10$ TeV, $\alpha=2$ is shown 
in Fig.\ref{ad_light_curve} (small panel). For magnetic fields we assume a
$B(r) \propto D^{-1}$ dependence with $B=0.1$~G at periastron.
{ To reconstruct the adiabatic loss profile, we first performed calculations with the "best guess" initial profile for adiabatic losses. Namely, if we assume that the  TeV lightcurve has a minimum around periastron, then obviously the adiabatic losses should increase closer to the periastron. Also, in order to avoid an overestimate of TeV fluxes compared to  the  fluxes (or upper limits) reported by HESS at phases well beyond periastron, one should assume some increase of adiabatic losses at large separations (see as well in \cite{kirk05}). Then we "improve" the shape of this profile using several iterations. }
  
It can be seen that in order to match the lightcurve shown in 
Fig. \ref{ad_light_curve} one needs a very sharp increase of 
adiabatic losses with characteristic time 
$T \sim 100$ sec. This can be naturally related 
to a much smaller size of the emission region at periastron, 
i.e. the region occupied by relativistic electrons 
accelerated by the termination shock is a denser region
closer to the star. In the case of termination of the 
wind in the highest density environment which coincides with the 
passage of the pulsar through the stellar disk we would expect 
even higher nonradiative losses, so one should expect 
some deviation from the adopted smooth symmetric profile of 
adiabatic losses shown in  Fig.\ref{ad_light_curve} (the small panel). 
Interestingly, the TeV gamma-ray flux at $t=20$ days from periastron is lower
in comparison with the adopted reference lightcurve. Note, that this "anomaly", related to the four points
after periastron passage, appears also in other models (see below). 
{ This "anomaly" can be interpreted as a result of the enhanced 
nonradiative losses in the disk. Another natural reason for the reduction of gamma-ray emissivity could be the deficit of the target photons in the stellar disk\footnote{ Even assuming that the photons from the star absorbed in the disk
are fully re-radiated at other wavelengths, one should nevertheless
expect significant reduction of the photon energy density in the disk 
because of the isotropisation of the 
initial radial distribution of photons from the star.}.}  However, the large  statistical 
and systematic errors of TeV fluxes do not allow certain conclusions 
in this regard. Therefore the reference lightcurve 
in Fig. \ref{ad_light_curve} should be treated as a reasonable approximation 
for derivation of basic parameters of the system.

\begin{figure}
\includegraphics[width=6cm,angle=270]{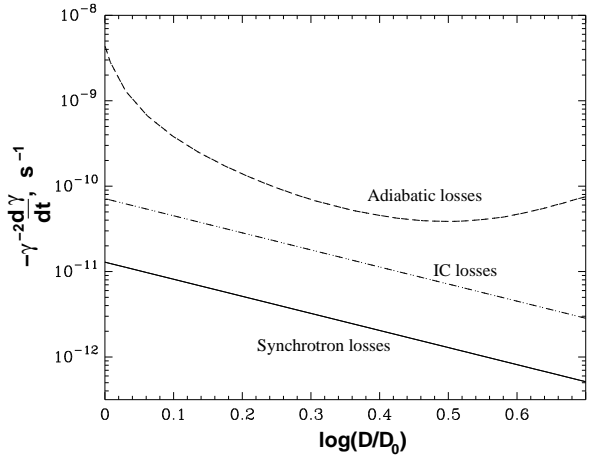}
\caption{The energy losses for 1 TeV electrons versus the 
separation distance between the pulsar and the companion star.  
The solid line is for synchrotron losses for $B=0.1$G at periastron, 
and $B(D) \propto D^{-1}$; the dashed-dotted line 
is for IC losses for black body distribution of target photons with temperature $T=2.3\cdot10^4$K diluted with coefficient $\kappa=(R_\star/2D)^2$, where $R_\star$ is the radius of the star. 
The dashed line corresponds to adiabatic losses for the reconstructed 
time-profile shown in Fig.\ref{ad_light_curve} (the small panel).}
\label{adiabat_sep}
\end{figure}

\begin{figure}
\includegraphics[width=8cm]{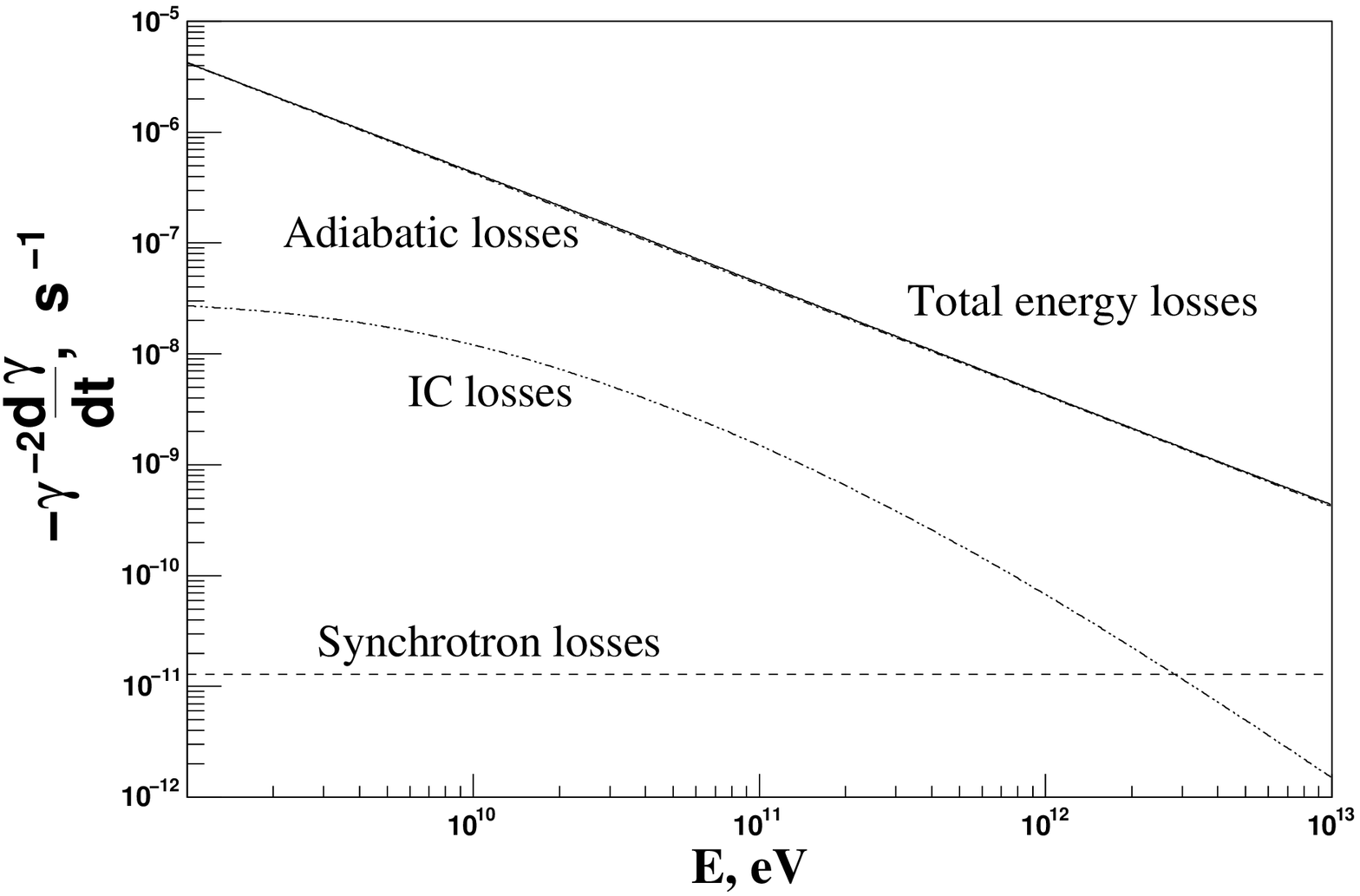}
\includegraphics[width=8cm]{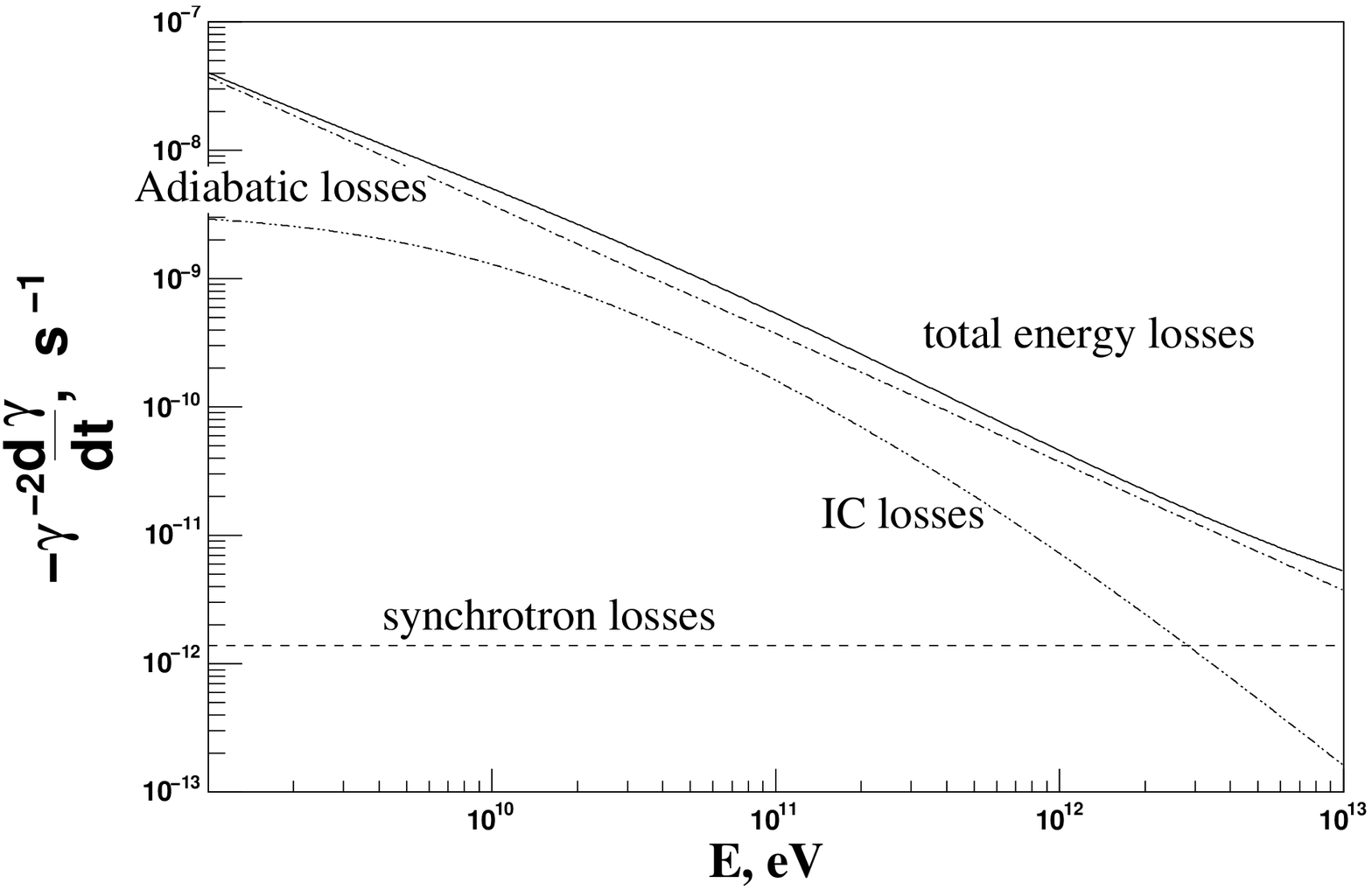}
\caption{The energy loss rates calculated for two 
epochs: at periastron (top panel) and at $t=35$ days
from periastron (bottom panel). The assumed parameters are 
the same as in Fig.\ref{adiabat_sep}.} 
\label{ad_maxmin}
\end{figure}

At epochs far from periastron the 
rate of the required nonradiative losses drops significantly, 
however it still remains faster than IC and synchrotron losses
with characteristic time $T \sim 10^4$ sec 
(see Fig.\ref{adiabat_sep}). Note that the curves 
in Fig.\ref{adiabat_sep} correspond to electrons of energy 1 TeV. 
However, as it is seen from Fig.\ref{ad_maxmin}, nonradiative 
losses should dominate over radiative losses at all energies of electrons and during the entire orbit of the pulsar.

\begin{figure}
\includegraphics[width=6cm,angle=270]{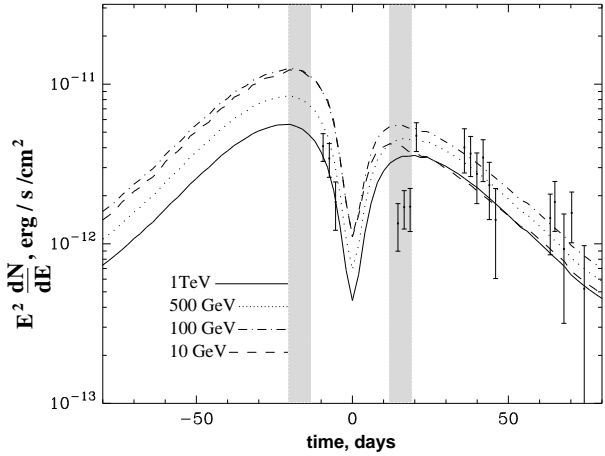}
\caption{Gamma-rays lightcurves in the nonradiative loss
dominated scenario are calculated for four different energies: 
$E_\gamma=1\,{\rm TeV}$ (solid line),   
$E_\gamma=0.5\,{\rm TeV}$ (dotted line),  
$E_\gamma=0.1\,{\rm TeV}$ (dash-dotted), 
$E_\gamma=10\,{\rm GeV}$ (dashed line). 
The HESS measurements \citep{aharonian05a} of 
1 TeV gamma-ray fluxes are also shown.  The model parameters are same as in Fig.\ref{adiabat_sep}.
}
\label{lightcurve2}
\end{figure}

\begin{figure}
\includegraphics[width=6cm,angle=270]{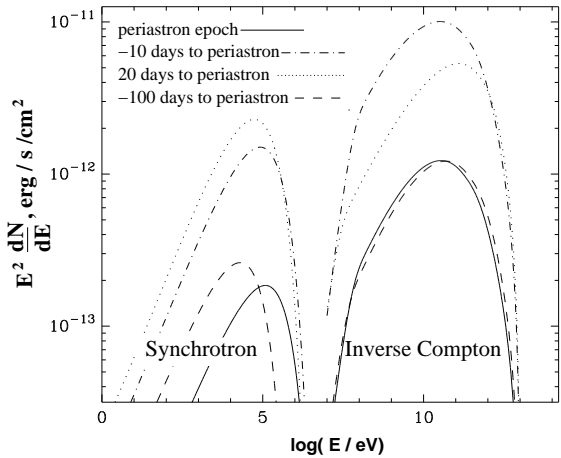}
\caption{Broadband spectral energy distribution of synchrotron and IC components of radiation in the nonradiative loss dominated scenario.
The spectra correspond  to $t=-100$ days (dashed), 
-10 days (dash-dotted), 0 days (solid), 
+20 days (dotted line) epochs (periastron is at $t=0$). 
The model parameters are same as in Fig.\ref{adiabat_sep}. 
}
\label{ad_rad}
\end{figure}

\begin{figure}
\includegraphics[width=8cm]{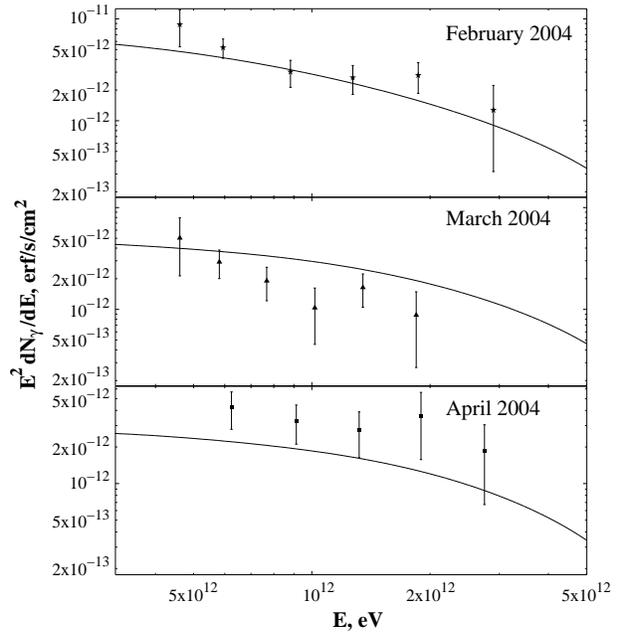}
\caption{The time-averaged (over the HESS observation periods)
TeV gamma-ray spectrum shown together 
with HESS measurements. 
The model parameters are same as in Fig.\ref{adiabat_sep}
}
\label{ad_spectra}
\end{figure}

In Fig.\ref{lightcurve2} we show lightcurves calculated for four different gamma-ray energies. The lightcurve for 1 TeV coincides, by definition, with the reference lightcurve shown in Fig.\ref{ad_light_curve}.
The lightcurves generally are similar what is explained by a dominance of adiabatic losses at all electron energies.  At the same time 
the energy spectra of IC radiation 
at different energy bands are significantly different. Indeed, 
the dominance of adiabatic or energy-independent escape losses
maintains the acceleration spectrum of electrons unchanged. 
Thus at energies $E \ll E_{\rm max}$, 
the gamma-rays produced in the Thompson regime ($E_\gamma \leq 10$ GeV) 
will have a power-law spectrum with photon index $(\alpha+1)/2$, 
while in the deep Klein-Nishina regime the spectrum will be 
proportional to $E_\gamma^{-(\alpha+1)} \ln E_\gamma$. 
This effect is seen in Fig.\ref{ad_rad}, where 
we show the broadband spectral energy distribution (SED)
of radiation at different epochs, 
consisting of synchrotron and IC components. 
{ In Fig. \ref{ad_spectra} we also show the gamma-ray 
spectra averaged over three periods of HESS observations 
in February, March and April 2004. Within the statistical 
and systematic uncertainties, the agreement with the 
fluxes  reported by HESS is satisfactory \citep{aharonian05a}.
} 

\begin{figure}
\includegraphics[width=4cm,angle=270]{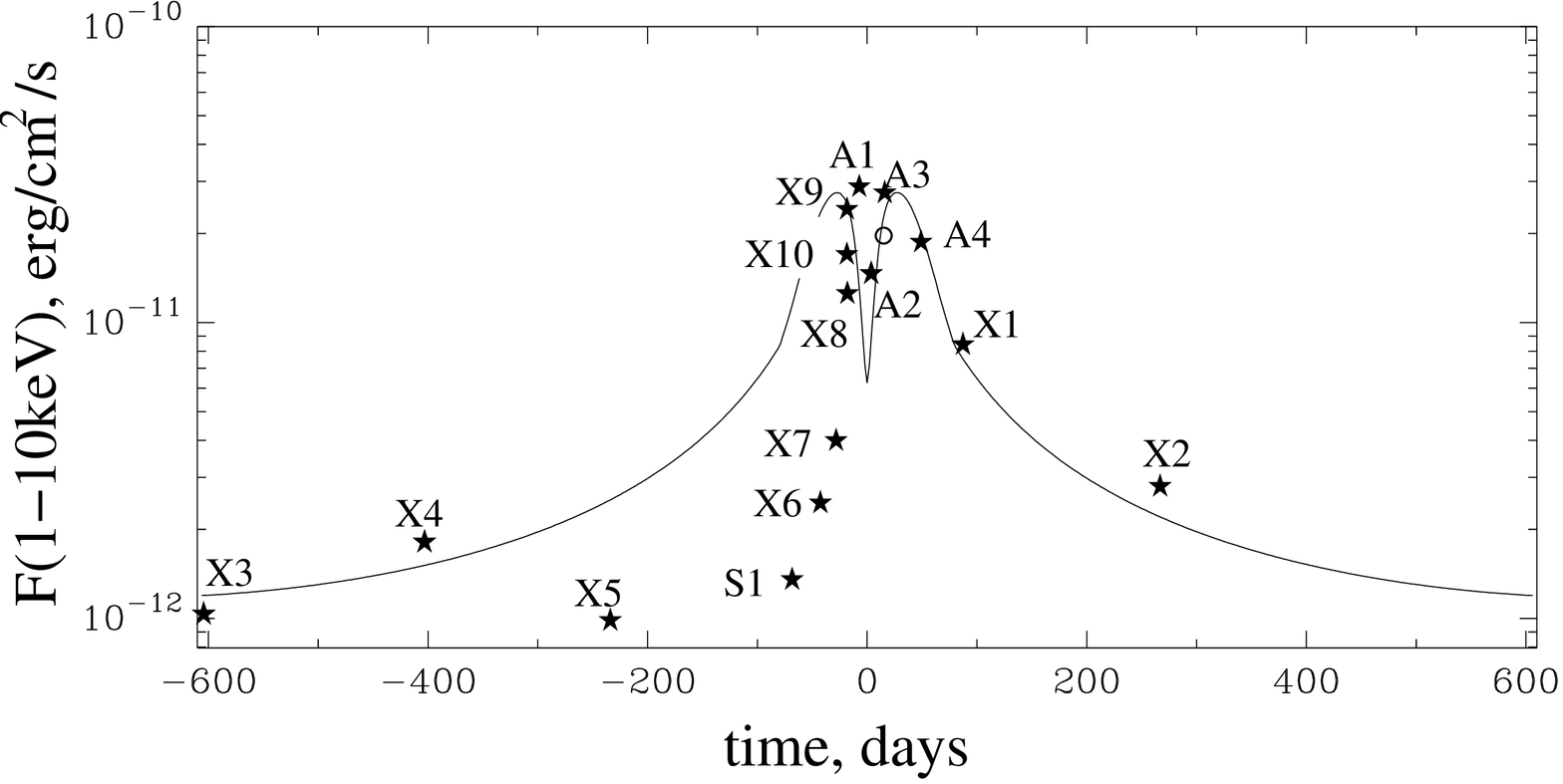}
\caption{The X-ray lightcurve calculated for the adiabatic loss time-profile shown in Fig.\ref{ad_light_curve}, and magnetic field strength $B=0.45(D/D_0)^{-1}$~G.
The point sets A1-A4, X1-X10, S1 correspond to ASCA \citep{hirayama96}, XMM-Newton \citep{chernyakova06} and BeppoSAX \citep{chernyakova06} observations, respectively. The open point corresponds to the 20-80 keV hard X-ray flux, $F_{\rm x} \sim 3\times10^{-11} \ \rm erg/cm^2 s$, as reported by the INTEGRAL team for the period 14.1-17.5 days after the periastron \citep{shaw04}.
}
\label{xla}
\end{figure}

{For the adopted key parameters, in particular for the
electron adiabatic loss rates shown in Figs. \ref{adiabat_sep}, \ref{ad_maxmin},  
the gamma-ray fluxes are not
sensitive to the  magnetic field strength as long as it 
does not exceed 0.5 G.  At the same time,  the synchrotron flux 
is very sensitive to the B-field ($\propto B^2$). It is seen in Fig.\ref{ad_rad}
that for the assumed magnetic field  0.1G the fluxes of 
synchrotron X-rays 
are significantly below $10^{-11} \rm \ erg/cm^2 s$, thus they cannot explain the 
X-ray data \citep{chernyakova06}. On the other hand, assuming 
a specific value of magnetic field  
$B_*=0.45$G at periastron, one can  increase the synchrotron X-ray flux 
to the observed level, and, at the same time keep the gamma-ray
fluxes practically unchanged. Not that any significant (20 \% or so) 
deviation from this value of the magnetic field 
would lead to the reduction of X-ray fluxes (for $B \le B_*$), or gamma-ray fluxes 
(for $B \ge  B_*$) below the observed flux levels. 
The X-ray lightcurve, calculated for the B-field 
dependence $B=0.45(D_0/D)$G, is shown  Fig.\ref{xla}. 
While the calculated lightcurve is in reasonable  
agreement with observations around the periastron and at high separations 
(more than 200 days), it significantly  exceeds  the 
fluxes between 200 days to several days before periastron.
This implies a much weaker magnetic field,
and thus the synchrotron radiation fails to explain the X-ray emission.  
A more speculative explanation could be that, the energy release in 
nonthermal particles for this period is suppressed, e.g. due to the 
the interaction with the stellar outflow. { If the X-ray and gamma-ray components are produced by the same electron population,}
this effect will have a similar impact on the gamma-ray lightcurve,
namely it should suppress significantly the gamma-ray fluxes. However
the lack of gamma-ray data for this period does not allow any conclusion 
in this regard. 
}


\section{Maximum energy of electrons}\label{maxenergy}
It is convenient to present the acceleration time of electrons 
in the following form:
\begin{equation}
t_{\rm acc}={\eta\, r_{\rm L}\over c}\approx 0.11\, 
E_{\rm TeV}\, B_{\rm G}^{-1}\, \eta \ {\rm s}\,,
\label{accel_time}
\end{equation}
where { $r_{\rm L}$ is the Larmor radius, and} $\eta$ is a dimensionless constant; 
$\eta=1$ corresponds to the maximum possible rate of acceleration 
allowed by classical electrodynamics. {{ It is well known theoretically, that in case of nonrelativistic parallel shocks  (with shock velocity $v$)}, $\eta>(c/v)^2\gg1$ (see e.g. \cite{pesses82,jokipii87,aharonian02}). { Although this parameter is not consistently constrained by theory,} $\eta$ can significantly exceed 1 even in the case of relativistic shocks. { In the case when the energy losses of electron are  dominated by 
synchrotron cooling, the maximum energy of the synchrotron radiation 
depends only on the $\eta$ parameter (see below for more details).
Thus the measurements of the synchrotron radiation in the cutoff region can give us quite robust information about $\eta$. However, this simple relation does not work in the case of PSR B1259-63, at least for the models discussed here which assume that electrons are cooled via inverse Compton scattering or due to adiabatic losses.  On the other hand, the spectral shape of fluxes of IC $\gamma$-rays, appear quite sensitive, especially at TeV energies, to the value of $\eta$. Thus if the suggested models describe the gamma-ray production scenarios correctly, we can derive information about $\eta$ from the comparison of model calculations with the observed gamma-ray spectra.}

In Fig.\ref{time1} we show characteristic acceleration times 
for 3 different values of $\eta=4 \times 10^3, \ 10^3, \ 10^2$,
together with synchrotron and Compton cooling 
timescales calculated for the epoch of the periastron 
assuming the magnetic field $B=0.05$~G. 
In  Fig.\ref{time1} the energy-independent escape time, which was assumed to be 
$10^4$~s, is also shown. 

The maximum energy of electrons is determined from the balance of particle acceleration and loss rates, in Fig.\ref{time1} this energy is defined by the intersection of curves corresponding to the acceleration
and loss times. Because of essentially different energy dependences of 
characteristic energy loss times 
$t_{\rm syn}$, $t_{\rm IC}$ and $t_{\rm esc}$, the maximum electron energy  is determined, depending on the value of 
$\eta$, by IC losses (a) or by escape (b) or by synchrotron losses (c)
(see Fig.\ref{time1}).  

If the electron energy losses are  
dominated by synchrotron cooling in the magnetic field $B_{\rm G}=B/1$~G
with characteristic time
\begin{equation}
t_{\rm syn} \approx 400 B_G^{-2} E_{\rm TeV}^{-1} \ {\rm s}\,,  
\end{equation}
the corresponding maximum energy of electrons is  
\begin{equation}
E_{\rm e,max}  \approx 2 \, B_{\rm G}^{-1/2} \, \left(\eta\over 10^3\right)^{-1/2}~{\rm TeV}\,.
\label{emaxsynch}
\end{equation}
Note  that in the case of $\eta={\rm const}$ the maximum energy of synchrotron photons 
does not depend on the magnetic field 
($E_{\rm syn,max}\propto E_{\rm e,max}^2\ B$=const), but depends on $\eta$, namely $E_{\rm syn,max}\sim 100(\eta/10^3)^{-1}\,{\rm keV}$. This relation
contains unique information about the acceleration rate through the $\eta$ parameter. 

\begin{figure}
\includegraphics[width=6cm,angle=270]{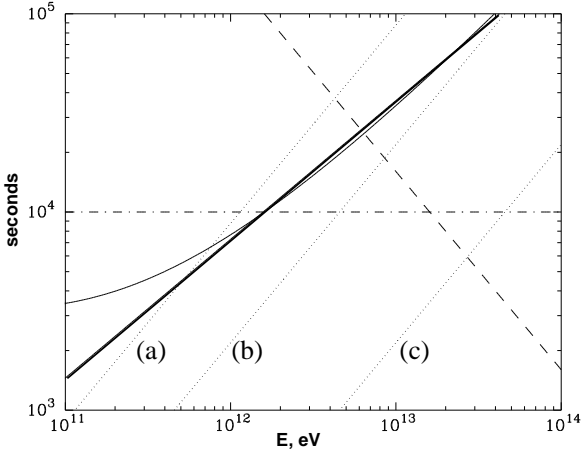}
\caption{The acceleration and cooling times of electrons at periastron. The solid line corresponds to the IC cooling time obtained with accurate numerical calculation; the thick solid line is the IC cooling time given by  the Eq. (\ref{ic_approxm}); the dashed line corresponds to the synchrotron cooling time ($B=0.05\ {\rm G}$); the dashed-dotted line is the escape time; the dotted lines are acceleration times for $\eta=4\cdot10^3$ (a),$\eta=10^3$ (b),$\eta=10^2$ (c).}
\label{time1}
\end{figure}

\begin{figure}
\includegraphics[width=6cm,angle=270]{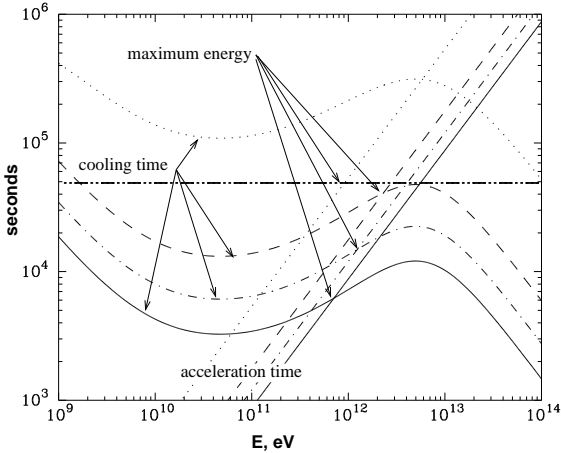}
\caption{The acceleration and energy loss times of electrons of different epochs. The combined IC + synchrotron cooling times are calculated assuming $B_0=0.05\left(D_{\rm 0}/D\right)\ {\rm G}$ and $T=2.3\cdot10^4\ {\rm K}$. The acceleration times are given by Eq.(\ref{accel_time}) assuming $\eta=4\cdot10^3$. The horizontal dash-dot-dotted line corresponds to the $T_{\rm esc}=5\cdot10^4{\rm s}$. The  solid lines correspond to the periastron epoch; the dash-dotted lines to $\pm10$ days; dashed lines to  $\pm20$days and dotted lines to $\pm100$days. The maximum  injection energy due to radiative cooling is determined by crossing of the same type lines. The maximum energy related to the electron escape is determined by crossing of an acceleration line with the escape line.}
\label{time}
\end{figure}

\begin{figure}
\includegraphics[width=3.5cm,angle=270]{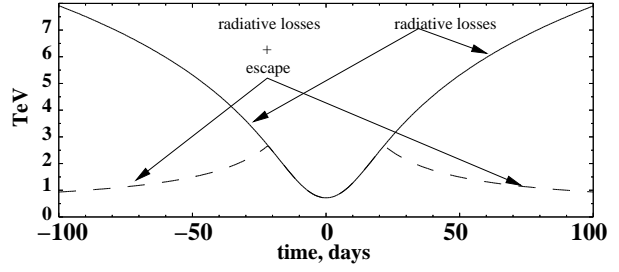}
\caption{Maximum energy of electrons at the pulsar wind termination shock. The solid line corresponds to the cutoff energy caused by radiative cooling. The dashed line corresponds to the case of combined radiative and escape losses. The model parameters are same as in Fig.\ref{time}.}
\label{cutoff}
\end{figure}

In the regime when IC cooling dominates over the synchrotron cooling, { the radiation cooling time is determined by}
\begin{equation}
t_{\rm IC} \approx 7\cdot10^3  
w_0^{-1} \ E_{\rm TeV}^{0.7}\ {\rm s}\, ,
\label{ic_approxm}
\end{equation}
where $w_0$ is the energy density of the target photons in $\rm erg/cm^{3}$ units.
In Fig.\ref{time1} we show the accurate numerical calculation of the
IC cooling time. It can be seen that above 1 TeV Eq.(\ref{ic_approxm}) provides quite accurate approximation of the IC cooling time.  

The corresponding maximum energy of accelerated electrons is 
\begin{equation}
E_{\rm e,max} \simeq  
9\cdot10^5 \ \left(B_{\rm G}/w_0\right)^{3.3}\left(\eta\over10^3\right)^{-3.3}\ 
{\rm TeV}\,.
\label{emaxic}
\end{equation}
This somewhat unusual  dependence of $E_{\rm e,max}$
on the photon density $w_0$ is the result of IC scattering
in deep  Klein-Nishina regime. Obviously, in the Thomson 
regime $E_{\rm e,max} \propto \left(B_{\rm G}/w_0\right)^{1/2} \eta^{-1/2}$. 
The very strong dependence of $E_{\rm e,max}$ in 
Eq.(\ref{emaxic}) on three highly variable parameters,
$B$, $w$ and $\eta$, allows variation of  $E_{\rm e,max}$
in very broad limits. For example, for the  $B \propto 1/D$
type dependence of the B-field, and assuming constant $\eta$, 
the increase of the separation between  the compact object
and the star by a factor of two would lead to the change of 
$E_{\rm e,max}$ by a factor of $2^{3.3} \simeq 10$, and 
correspondingly to dramatic variation of the flux of highest energy gamma-rays. 

Finally, the escape of electrons may also have a strong impact on the 
variation of $E_{\rm e,max}$ depending on the position of the pulsar. { Actually the effective escape of electrons from the acceleration site is somewhat shorter than the escape time. So in this case, ignoring the radiative energy losses of electrons and taking the upper limit for the escape from accelerator, one has }
%
%
\begin{equation}
E_{\rm e,max} \simeq  9 \,B_{\rm G} \left(T_{\rm esc}\over10^3 {\rm s}\right)\left(\eta\over10^3\right)^{-1} \ {\rm TeV}\,,
\label{emaxesc}
\end{equation}
where $T$ is the escape time of electrons.

Obviously, all relevant timescales depend on the pulsar position
in the orbit, therefore the high energy cutoff in the spectrum of electrons is expected to be variable. { As long as there no theoretical predictions for possible $\eta$ parameter dependence on physical conditions in the accelerator, in what follows we assume $\eta$ to be constant.} In Fig.\ref{time} we show the radiation and acceleration timescales for different epochs -- at periastron and  $\pm10,20,100$ days 
from the periastron. For the chosen model parameters, $B=0.05(D_{\rm 0}/D)$~G and 
$\eta=4 \times 10^3$, the cutoff in the electron spectrum at the periastron is determined by IC losses, while at large separation distances the 
synchrotron and escape losses play the more important role in formation of the cutoff.
This is demonstrated in  Fig.\ref{cutoff}, where the high energy cutoff in the electron spectrum is shown as a function of epoch. Solid line corresponds to
the case of radiation (IC and synchrotron) losses. In this case one expects 
significant reduction of the cutoff energy at epochs close to the  
periastron, where strong IC losses push the cutoff energy down to 
$\leq 1$~TeV. Far from the periastron,  the cutoff energy can increase
up to 10 TeV, unless the losses due to escape become dominant. 
The IC cooling time at the epoch with separation $D$ 
is $t_{\rm cool}\simeq10^3{\rm s} (D/D_0)^2$~s. Therefore, if the
characteristic escape time is about $10^4$~s, the impact of particle 
escape becomes important for 
separations $D \geq 3 D_0$.  This effect is demonstrated in Fig.\ref{cutoff}
where (time and energy-independent) escape time 
$T_{\rm esc}=5 \times 10^4$~s is assumed.  One can see that for chosen 
model parameters the cutoff energy is a weak function of time with 
a local minimum ($\simeq 0.5$~TeV) at periastron, and
two maxima ($\simeq 2.5$~TeV) at $\pm 20$ days.     

It is important to note that the introduction of escape losses is crucial for 
explanation of the observed TeV lightcurve in this scenario.
Indeed, while the reduction of the cutoff energy 
in the spectrum of electrons due to enhanced IC losses 
satisfactorily explains the minimum at the periastron, { this would imply much} higher fluxes 
at large separations in contrast to the HESS observations. The additional assumption that
electrons suffer also significant escape losses ($T_{\rm esc}=5 \times 10^4$~s) 
allows dramatic suppression of  the gamma-ray fluxes beyond $|t| > 20$ days (compare dashed and dot-dashed curves in 
Fig.\ref{lightcurve1}).

The impact of the variation  of relative contributions of radiative and 
escape losses on the formation of the energy distributions of electrons is demonstrated in Fig.\ref{density}. The corresponding lightcurves of 
inverse Compton gamma-rays at $E_\gamma=1$ TeV, 500 GeV, 100 GeV and 10 GeV, and 1-10 keV synchrotron photons  are shown in Fig.\ref{cutoff_lightcurve}. For comparison the HESS measurements \citep{aharonian05a} of 1 TeV gamma-ray fluxes are also shown.
The agreement of calculations with the HESS lightcurve is rather satisfactory except for somewhat higher predicted flux at the epoch of 2 weeks after the periastron which coincides with the pulsar passage through the stellar disk. In Fig.\ref{cutoff_spectra} we compare the energy spectrum reported by HESS with the average TeV gamma-ray 
spectrum calculated for the period of the HESS observations in February 2004. Although it  is possible to achieve a better agreement with the measurements, at this stage the attempt for a better spectral
fit could be hardly justified given the  statistical and systematic uncertainties of the measurements.

Through a variation of $E_{\rm e,max}$ the lightcurves at TeV and GeV energies can have quite different profiles. 
Namely, the TeV lightcurves have a clear minimum at periastron 
which is explained by the sub-TeV cutoff in the spectrum of accelerated electrons. At the same time this cutoff in the electron spectrum is still sufficiently high and therefore does not have a strong impact at GeV energies. Therefore the GeV lightcurves show their maximum a few days before the  periastron \footnote{Note that the shift of the position of the maximum is caused, as discussed above, by the anisotropy of the Compton scattering, but not by the change of the target photon density 
as long as the IC proceeds in the "saturation regime".}.
It is important to note that the significant drop of 
gamma-ray fluxes at large separations is due to the escape losses, otherwise one should expect rather constant flux with a weak maximum close to periastron.  Also this model predicts different energy spectra of gamma-ray below 100 GeV at different epochs. Indeed, at large separations, when the escape losses dominate, the injection spectrum 
of electrons remains unchanged, therefore we expect noticeably harder gamma-ray spectra
in the GeV energy band at epochs $|t| > 20$~days (see  Fig.\ref{cutoff_rad}).

Remarkably, the calculated fluxes at GeV energies are well above the sensitivity of 
GLAST which makes this source a perfect target for future observations with GLAST.
It should be noted, however, that the fluxes at GeV energies could be significantly suppressed
due to a possible low energy
cutoff in the acceleration spectrum of electrons.{ This is a standard assumption in the models of PWN, see e.g. \cite{kennel84,kirk99}.}    

In this scenario,  the magnetic field energy density at the shock should be significantly
less than the energy density of stellar photons. Thus, assuming 
the same strength of the magnetic field in the acceleration and
radiation regions, one obtains quite low synchrotron 
fluxes (see Fig.~\ref{cutoff_lightcurve}). This would imply that the observed X-rays have 
non-synchrotron origin (e.g. IC origin; see  \citep{chernyakova06}). 
Another possibility is to assume that the magnetic field in the radiation region
is somewhat higher than at the shock (note that that a similar situation 
takes place in the Crab nebula \citep{kennel84}). 
In Fig.\ref{xlc} we show the lightcurve of 1-10 keV X-rays, 
assuming  that in the radiation region the magnetic field is stronger  by 
a factor of 8. If so, the synchrotron X-ray flux could achieve 
the observed flux level. In this scenario the X-ray and gamma-ray production 
regions are essentially different, although they could partly overlap. 
While the bulk of X-rays is formed in a  magnetized 
region(s) far from the shock (where the electrons are accelerated), 
the IC gamma-rays come from more extended zones, which include also 
the site of particle acceleration. { In Fig.\ref{xlc} we show also the 
20-80 keV hard X-ray flux as reported by the 
INTEGRAL team for the 14.1-17.5 days after the periastron \cite{shaw04}. 
Note that these days correspond to the minimum of the gamma-ray fluxes which 
we explain quantitatively by the passage of the pulsar through the disk, e.g. due to the 
the deficit of the target radiation field. 
}

\begin{figure}
\includegraphics[width=6cm,angle=270]{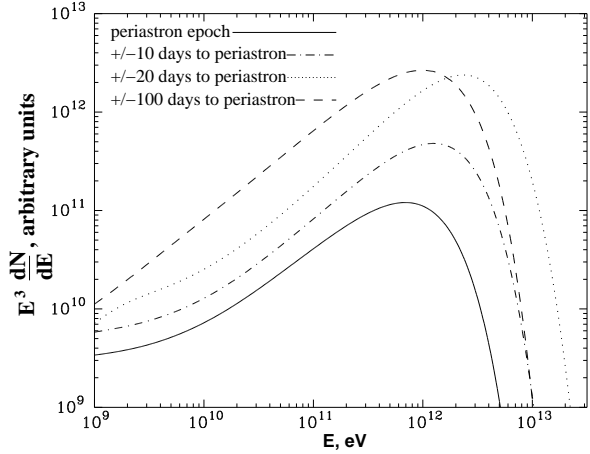}
\caption{Electron energy distributions at different epochs ($0,\,\pm10,\,\pm20,\,\pm100\,{\rm days}$ to periastron passage). The model parameters are same as in Fig.\ref{time}.}
\label{density}
\end{figure}

\begin{figure}
\includegraphics[width=6cm,angle=270]{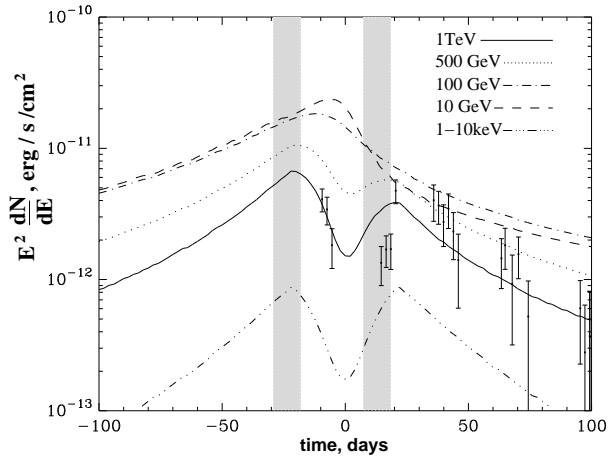}
\caption{ The gamma-ray lightcurves expected in the scenario of variation of the energy cutoff. The solid line: $E_\gamma=1\,{\rm TeV}$,  the dotted line: $E_\gamma=0.5\,{\rm TeV}$, the dash-dotted line: $E_\gamma=0.1\,{\rm TeV}$, the dashed line: $E_\gamma=10\,{\rm GeV}$ and the dash-dot-dot-dotted line: $E_\gamma=1-10\,{\rm keV}$ (synchrotron). The model parameters are same as in Fig.\ref{time}. The injection spectrum was assumed to be $Q(E_{\rm e})\propto E_{\rm e}^{-2}{\rm exp}\left(-E_{\rm e}/E_{\rm e,max}\right)$. The injection acceleration rate of electrons was assumed to be at level of 5\% of the pulsar spindown luminosity. The HESS measurements \citep{aharonian05a} of 
1 TeV gamma-ray fluxes are also shown. The vertical shadowed zones correspond to the stellar disk location.}
\label{cutoff_lightcurve}
\end{figure}
\begin{figure}
\includegraphics[width=6cm,angle=270]{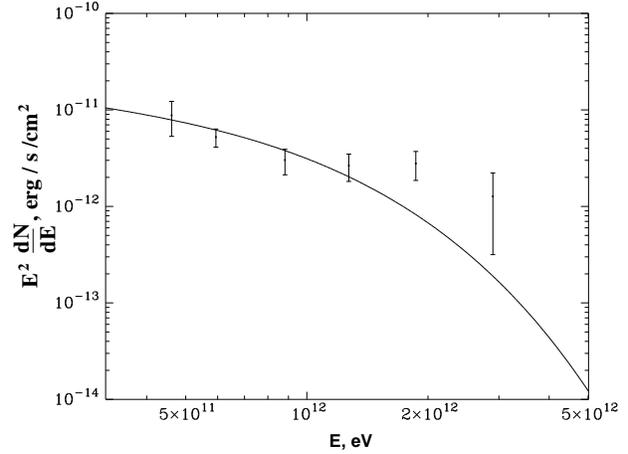}
\caption{The averaged spectrum of IC gamma-rays compeared with the spectrum measured by HESS during the period Feb. 2004. The model parameters are same as in Fig.\ref{cutoff_lightcurve}.}
\label{cutoff_spectra}
\end{figure}

\begin{figure}
\includegraphics[width=6cm,angle=270]{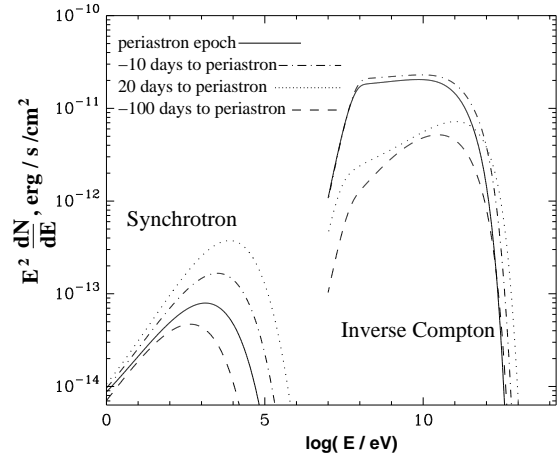}
\caption{Gamma-ray spectra of synchrotron and IC radiation at -100 (the dashed line),-10 (the dotted-dashed line),0 (the solid line),+20 (the dotted line) days relative to the periastron passage.  The model parameters are the same as in Fig.\ref{cutoff_lightcurve}.}
\label{cutoff_rad}
\end{figure}
\begin{figure}
\includegraphics[width=4cm,angle=270]{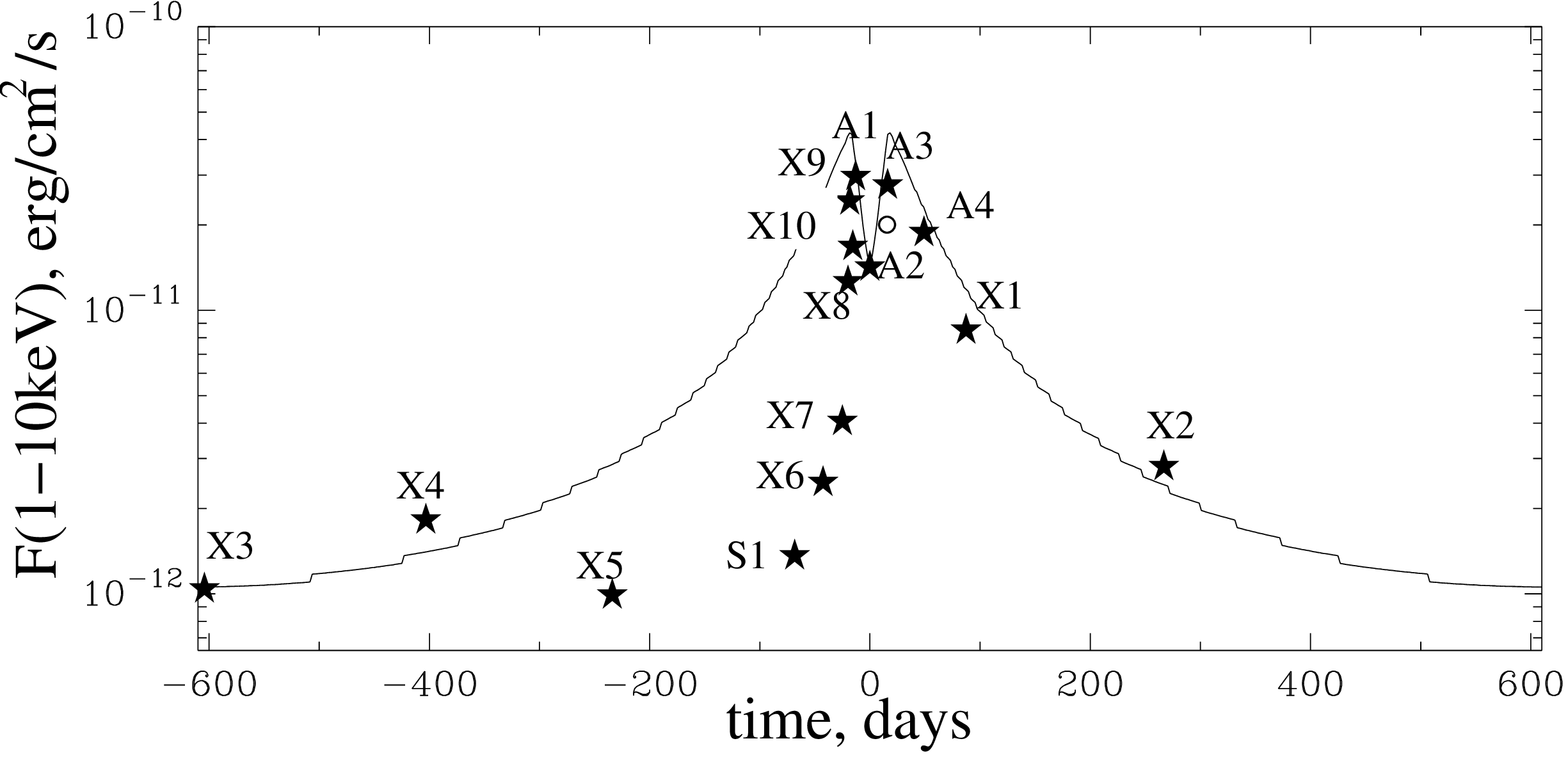}
\caption{The X-ray lightcurve calculated for the scenario with early sub TeV cutoff. At shock front the magnetic field was assumed to be $B_0=0.03\left(D_{\rm 0}/D\right)\ {\rm G}$, and in the emitting region to be higher by a factor of 10. The acceleration efficiency has value $\eta=4\times10^3$. An orbital-dependent escape time was assumed to be $t_{\rm esc}=1.75\times10^4\left(D/D_{\rm 0}\right)^{1/2}$~s.
The point sets A1-A4, X1-X10, S1 correspond to ASCA \citep{hirayama96}, XMM-Newton \citep{chernyakova06} and BeppoSAX \citep{chernyakova06} observations, respectively. The open point corresponds to the 20-80 keV hard X-ray flux, $F_{\rm x} \sim 3\times10^{-11} \ \rm erg/cm^2 s$, as reported by the INTEGRAL team for the period 14.1-17.5 days after the periastron \citep{shaw04}.
}
\label{xlc}
\end{figure}

\section{Comptonization of the unshocked wind}\label{wind}

While in the previous sections we tried to explain 
the observed modulation of TeV gamma-rays fluxes 
by energy losses of accelerated electrons at the 
termination of the wind, it is interesting to investigate 
whether one can relate the observed TeV lightcurve to   
the Compton losses of the kinetic energy of the bulk motion 
of the cold ultrarelativistic wind. 
 
Generally this effect can be important in a binary system with  
a high luminosity companion star. 
Although the electrons in cold  pulsar winds do not suffer 
synchrotron losses, a significant fraction of original kinetic energy of the bulk motion of the  electron-positron wind 
could be radiated away due to Comptonization. Thus  
the power available for acceleration of electrons 
at termination of the wind depends on the position of the pulsar. Obviously, 
in this scenario we expect a minimum flux of gamma-rays closer to periastron.  In other words, while in the previous section 
the modulation of the gamma-ray flux is linked to the $E_{\rm e,max}$,
in this scenario the gamma-ray flux variation depends on the parameter $A$ characterizing the acceleration power of electrons given by Eq.(\ref{injection}).
  
According to the standard PWN model \citep{kennel84} 
the { isotropic} cold electron-positron wind\footnote{ Here we follow a model suggested by \cite{kennel84}, assuming an isotropic pulsar wind, but it is worthy to note that the pulsar wind can be strongly anisotropic \citep{bogovalov02}, thus the non-typical lightcurve can be a result of the interaction of two anisotropic winds.} has a typical 
bulk motion Lorentz factor $\Gamma\propto10^{4}-10^{6}$, thus the interaction 
of the wind electrons with starlight in the Klein-Nishina limit should 
lead to the formation of 
a narrow gamma-ray component with typical energy $\Gamma m_ec^2$. 
This effect also leads to the modulation of the bulk motion Lorentz factor as shown in Fig.\ref{deacy}. The calculations in Fig.\ref{deacy} were performed for $L_{\rm star}=2.2\cdot10^{38}{\rm erg/s}$ and different values of the initial Lorentz factor of the pulsar wind. Note that in the Thomson regime $t_{\rm cool} \propto \Gamma^{-1}$, i.e. the decrease of wind Lorentz factor  leads to the increase of the cooling time. On the other hand, in the Klein-Nishina regime  the cooling time
increases with Lorentz factor as $t_{\rm cool} \propto  \Gamma^{0.7}$. 
Therefore the maximum effect is 
achieved in the Thomson-to-Klein-Nishina transition region, i.e.  
around $\Gamma=10^5$. 

As it follows from  Fig.\ref{deacy}, the initial Lorentz factor of 
the wind for $\Gamma_{\rm 0}=10^5$ is reduced by $40 \%$ at periastron. 
Interestingly, minimum 
reduction of the initial Lorentz factor ($ \sim 5 \%$) happens around  $t=\pm 20$ days (see in Fig.\ref{deacy}). Since the kinetic energy of the wind radiated away due to the Comptonization is determined by the starlight density and the length of the unshocked wind $\Delta l=r_{\rm sh}$, the  lightcurve is explained by the combination of two factors -- dependence of the starlight density on the separation $D$, and the distance to the wind termination point (it is assumed that the gas density of the stellar wind 
decreases as $D^2$). Obviously, in the case of electron-positron pulsar wind, the kinetic energy of bulk motion of the wind, and consequently the rate of shock accelerated electrons  
have  similar time behaviors: $A=A_{\rm 0}\Gamma(t)/\Gamma_{\rm 0}$, where $A_{\rm 0}$ 
characterizes the original power of the wind. 

Although qualitatively this behavior agrees with the TeV lightcurve 
detected by HESS, the effect of reduction of the kinetic energy 
of the wind is not sufficient to explain quantitatively the observed TeV lightcurve. Indeed, the Comptonization of the wind can lead to the 
reduction of the energy flux of TeV gamma-rays at 
periastron by only a factor of $\leq 1.5$, while 
the HESS observations show more significant
variation of the gamma-ray flux. Assuming somewhat
larger, by a factor of two, luminosity of the optical star, one
can get a  better agreement 
with the observed TeV lightcurve.  However, the range of luminosity of the star discussed in the literature favors a lower luminosity of the star \citep{tavani97,kirk99}. Therefore the effect of Comptonization of the pulsar wind cannot
play, even for $\Gamma_{\rm 0} \sim 10^5$, 
a major role in the formation of the TeV lightcurve. 

Even so, this effect cannot be ignored in the calculation of  the overall gamma-radiation of the system. Namely the Comptonization of the ultrarelativistic
pulsar wind unavoidably leads to an additional component of gamma-rays
produced at the pretermination stage of the wind. Due to the inverse Compton scattering of monoenergetic electrons on target photons with narrow, e.g. Planckian distribution,  
we should expect a specific, especially for $\Gamma_{\rm 0} \geq 10^5$, line-type gamma-ray emission { \citep{bogovalov00,kirk00,ball01}. \cite{kirk00} have calculated the IC radiation of the freely expanding wind, i.e. under assumption that the wind was not terminated. These calculations can be hardly applied for this system, since we see broad-band IC radiation of the shocked wind. As it was shown in the paper by \cite{ball01}, such an assumption leads to significant (an order of magnitude) overestimation of the gamma-ray flux. Unfortunately in the paper by \cite{ball01} there are only calculations for one value of Lorentz factor $\Gamma=10^6$. To supply this gap we performed calculations for a number of probable Lorentz factors.} The results of calculations of gamma-ray spectra of the unshocked wind are shown in Fig.\ref{fig7}. 
Comparison of these calculations with the average energy spectrum of PSR~B1259-63
measured by HESS excludes the initial Lorentz factor of the wind $\Gamma_{\rm 0} = 10^6$. Otherwise, the flux of the Comptonized emission of the unshocked wind would exceed the { observable} flux  even at apastron (see Fig.\ref{fig7}).  Due to the energy range  of gamma-rays  from this source available for HESS ($E \geq 300$~GeV), future observations unfortunately cannot significantly improve this limit. 
On the other hand, such studies can be performed by GLAST the 
sensitivity of which seems to be adequate, as is shown Fig.\ref{fig7},
for a deep probe of the initial wind Lorentz factors 
of PSR B1259-63 within $10^4$ to $10^6$. 
Thus, GLAST has a unique potential to prove the current pulsar wind paradigm
which assumes that the bulk of the spin-down luminosity  of the pulsar is
transformed to a cold wind  with Lorentz factor  exceeding $10^4$.

\section{Summary}
One of the recent exciting results of observational gamma-ray astronomy is 
the detection of TeV gamma-ray signal from the binary system PSR B1259-63/SS2883
\citep{aharonian05a}. 
While the absolute fluxes and energy spectra of TeV emission detected by HESS can be explained 
quite well in the framework of inverse Compton model \citep{kirk99}, the observed TeV 
lightcurve appears to be  significantly different from early predictions. This { can be considered as an argument in favor of} 
alternative (hadronic) models which relate the maximum in the TeV lightcurve observed 
after approximately three  weeks of the periastron to the interaction of the 
pulsar wind with the stellar disk \citep{kawachi04,chernyakova06}. { However, the hadronic models require a rather specific location of the stellar disk, which needs to be confirmed, since this assumption does not agree with the location of the stellar disk derived from the pulsed radio emission observations \citep{bogomazov05}}.    
   
{ The main objective of this paper is to show that the HESS results can be 
explained, under certain reasonable  assumptions concerning the cooling of
relativistic electrons, by  inverse Compton scenarios of  gamma-ray 
production in PSR B1259-63/SS2883. Namely, we study  three different scenarios of formation of gamma-ray lightcurve 
in the binary system PSR B1259-63/SS2883 with an aim 
to explore whether one can explain the observed TeV lightcurve 
by electrons accelerated at  the pulsar wind 
termination shock.} The natural target for the inverse Compton scattering in such a system 
is the thermal radiation from the optical star. 
Since the basic parameters characterizing the system are well known,
the predictions of gamma-ray fluxes at different epochs can be 
reduced to a calculation 
of the energy distribution  of the relativistic electrons 
under certain assumptions concerning the 
acceleration spectrum of electrons and of both their  
nonradiative (adiabatic and escape) and radiative (Compton and synchrotron) energy loss origin.
In particular, we demonstrate that the observed TeV lightcurve can be explained
(i) by adiabatic losses which dominate over the entire trajectory of 
the pulsar with a significant increase towards periastron, or 
(ii) by the variation of the cutoff energy in the acceleration spectrum of electrons 
due to the modulation of rate of inverse Compton losses depending on the position of the pulsar 
relative to the companion star. { Our models failed to explain the first four data points obtained just after periastron, what can be explained by a possible interaction with the Be star disk, which introduces additional physics not included in the presented model.}  Although we deal with a very complex  system,
we demonstrate  that the observed TeV lightcurve can be naturally explained
by the inverse Compton model under certain physically well justified 
assumptions.  Unfortunately, the large systematic 
and statistical uncertainties, as well as the relatively narrow energy band  of
the available TeV data do not allow robust constraints on several key model parameters like the magnetic field, escape time, acceleration efficiency, \textit{etc}. 
This also does not allow us to distinguish between 
different scenarios discussed above. 
In this regard, the future detailed 
observations both in MeV/GeV and TeV bands by GLAST and HESS
closer to periastron, as well as at the epochs  
when the pulsar crosses the stellar disk, will provide 
strong insight into the nature of this 
enigmatic object. Equally important are the detailed observations 
of  X-rays, e.g.  with  Chandra, XMM and Suzaku telescopes. The analysis
of gamma and X-ray data obtained simultaneously  should allow 
extraction of several key parameters characterizing the binary system.

Finally, although the Compton cooling   
of the unshocked electron-positron wind does  
have significant impact on the formation of the TeV lightcurve,  
the specific, line type gamma-radiation caused 
by the Comptonization of the cold ultrarelativistic 
wind should unavoidably appear either at GeV or TeV energies  
depending on the initial Lorentz factor  of the wind. 
Detection of this component of gamma-radiation, in particular by GLAST, 
of the unshocked wind 
will  provide unique information on  the formation and dynamics 
of pulsar winds. 

\section*{Acknowledgments}
We are grateful to Valenti Bosch-Ramon and Andrew Taylor for their useful comments and help with preparation of the manuscript.

\begin{figure}
\includegraphics[width=6cm,angle=270]{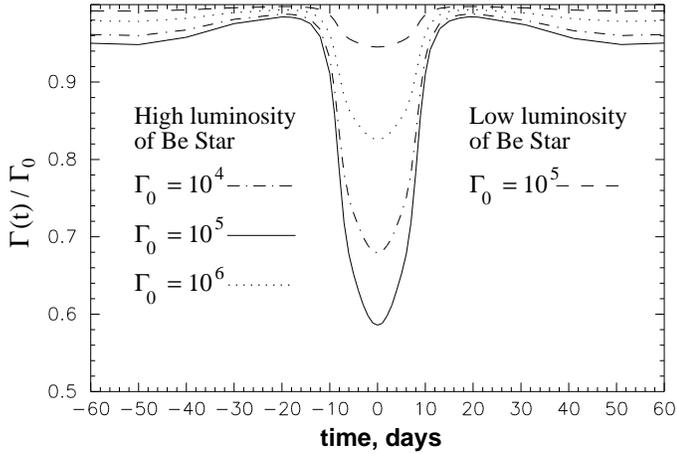}
\caption{The time-evolution of the wind Lorentz factor just before the termination.  The results are presented for the initial Lorentz factors 
$\Gamma_{\rm 0}=10^5$ (solid line), $\Gamma_{\rm 0}=10^6$ (dotted line)
and for $\Gamma_{\rm 0}=10^4$ (dashed-dotted line) for the luminosity of the companion star $ L_{\rm star}=2.2\cdot10^{38}$~erg/s. Dashed line shows the time evolution of the wind Lorentz factor for $\Gamma_{\rm 0}=10^5$  calculated the luminosity of the companion star $ L_{\rm star}=3.3\cdot10^{37}$~erg/s.}
\label{deacy}
\end{figure}

\begin{figure}
\includegraphics[width=6cm,angle=270.0]{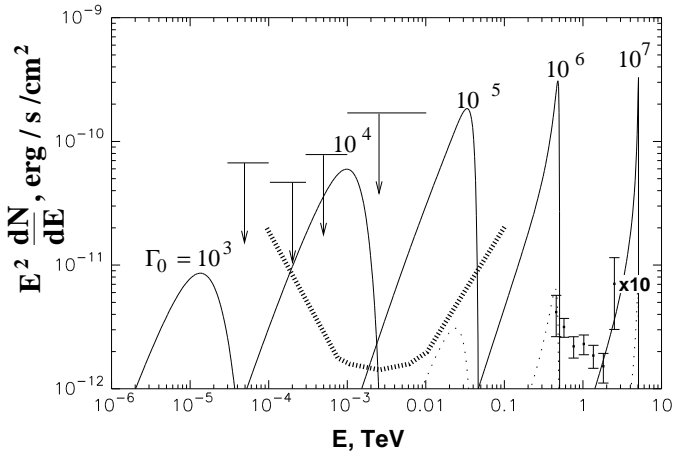}
\caption{The energy spectra of gamma-rays due to the Comptonization of the unshocked wind at the periastron (solid lines) and apastron (dotted lines) are calculated for initial Lorentz factor $\Gamma_{\rm 0}=10^3, 10^4,10^5,10^6$ and $10^7$ (indicated at the curves). The luminosity of the companion star was assumed $L_{\rm star}=2.2\cdot10^{38}$~erg/s. The experimental points correspond to the average energy spectrum measured by HESS within several weeks around the periastron, note that the last measurement was multiplied by a factor of 10. The upper limits in energy range 30MeV-3GeV correspond to the EGRET measurements \citep{tavani96}.  The differential flux sensitivity of GLAST for a point source is also shown. The corresponding curve represents sensitivity for one-year all-sky survey taken from http://www-glast.slac.stanford.edu/software/IS/glast\_lat\_performance.htm. However, since gamma-ray fluxes can be effectively observed only at the epochs not far from the periastron, the typical available observation time by GLAST would be limited by $\Delta t\leq 3$~weeks. This implies that the minimum detectable gamma-ray flux of GLAST shown in the figure should be increased by a factor between $(1{\rm year}/\Delta t)^{1/2}\simeq4$ and $16$ depending whether  the sensitivity is determined by the background or by the gamma-ray photon statistics.  
}
\label{fig7}
\end{figure}

\label{lastpage}
\end{document}